\documentclass[prl,a4paper,twocolumn,showpacs,amsmath]{revtex4}
\usepackage{graphicx}
\begin{document}
\title{Partonic Pole Matrix Elements for Fragmentation}
\author{S.~Mei{\ss}ner}
\affiliation{Institut f{\"u}r Theoretische Physik II,
 Ruhr-Universit{\"a}t Bochum, 44780 Bochum, Germany}
\author{A.~Metz}
\affiliation{Department of Physics, Temple University,
 Philadelphia, Pennsylvania 19122-6082, USA}
\date{5 May 2009}
\begin{abstract}
A model-independent analysis of collinear three-parton correlation functions
for fragmentation is performed.
By investigating their support properties it is shown, in particular, that the
so-called partonic pole matrix elements vanish.
This sheds new light on the understanding of transverse single spin
asymmetries in various hard semi-inclusive reactions.
Moreover, it gives additional strong evidence for the universality of
transverse-momentum-dependent fragmentation functions.
\end{abstract}
\pacs{12.38.Bx, 12.39.St, 13.85.Ni, 13.87.Fh}
\maketitle

%
%
{\it I. Introduction and definitions.}\,---\,QCD factorization theorems separate
the cross section for a large number of
hard processes into a perturbatively calculable part as well as
nonperturbative correlation functions~\cite{collins_89}.
These correlation functions are parton distributions (PDFs), fragmentation
functions (FFs), etc.~\cite{collins_81c,jaffe_91b}.
So far the main focus was on the leading (twist-2) two-parton correlators, and
in the meantime an enormous amount of information about these objects has been
obtained.

However, there is a set of interesting phenomena which\,---\,in the typically
used collinear picture with partons moving in the direction of their parent
hadrons\,---\,cannot be described by means of twist-2 correlators.
Perhaps the most intriguing observables among those are transverse single spin
asymmetries (SSAs) for which effects of up to $40 \,\%$ were observed in hard
hadronic reactions such as $p^{\uparrow} p \to \pi X$ and
$\bar{p}^{\uparrow} p \to \pi X$ at FermiLab~\cite{adams_91a,bravar_96} and
more recently also at the Relativistic Heavy Ion
Collider~\cite{adams_03,adler_05,arsene_08,abelev_08}.
In a factorized QCD approach one can generate such asymmetries by using
particular collinear twist-3 three-parton correlators.
This idea was brought up in~\cite{efremov_81} and worked out in more detail
in~\cite{qiu_91a} (see also Refs.~\cite{kouvaris_06,eguchi_06b}).
If one of the three partons becomes soft, one hits the pole of a particle
propagator in the hard part of a process providing an imaginary part
(a nontrivial phase) which, quite generally, can result in
SSAs~\cite{efremov_81,qiu_91a}.
Therefore, multiparton correlators with one (or more) soft partons are called
gluonic pole or fermionic pole matrix elements, depending on whether the soft
parton is a gluon or a quark.

Already a lot of work on different aspects of these partonic pole matrix
elements (PPMEs) exists, which has considerably improved our understanding
about such objects and the physics underlying SSAs.
So far the papers in the field of PPMEs were mostly dealing with correlators
that describe the parton structure of the target (PPMEs for parton
distributions).
The knowledge about multiparton correlators for fragmentation, however, is
still rather poor, and it is the aim of the present Letter to give new insights
into these objects.

Nonzero PPMEs for fragmentation would contribute to a large number of
transverse SSAs.
By using simple models for such three-parton matrix elements pioneering
phenomenological studies on the processes $p^{\uparrow} p \to \pi X$,
$p p \to \Lambda^{\uparrow} X$, $e p^{\uparrow} \to e \pi X$, and
$e p \to e \Lambda^{\uparrow} X$ were carried out
in~\cite{koike_01,eguchi_06a}.
Other work discussed PPMEs for fragmentation in connection with
$p^{\uparrow} p \to \pi \pi X$,
$p^{\uparrow} p \to \textrm{jet} \, \pi X$~\cite{bacchetta_05,bomhof_06}, as
well as $p p \to (\Lambda^{\uparrow} \textrm{jet}) \, \textrm{jet} \, X$ with
a $\Lambda$ hyperon detected in a jet~\cite{boer_07a}.
Note that this already long list of semi-inclusive reactions,
which might be related to PPMEs for fragmentation, is not yet complete.

Despite the potential importance of PPMEs for fragmentation it was, however,
so far not even clear whether they exist.
In fact, recent spectator model calculations imply (show) that these
objects vanish~\cite{metz_02,collins_04,gamberg_08}.
In this Letter it is proven for the first time in a fully model-independent
way that PPMEs for fragmentation vanish, which constitutes the main outcome of
our work.
To obtain this result we analyze the support properties of collinear twist-3
three-parton fragmentation correlators.

\begin{figure}[t]
 \includegraphics{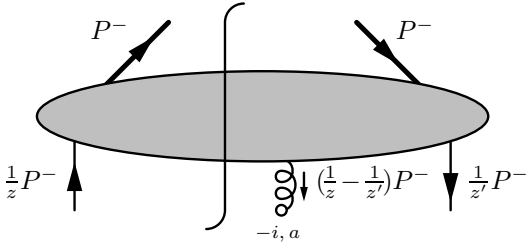}
 \caption{Kinematics of the $qgq$ correlator in Eq.~(\ref{e:qgq}).
 The wiggly line with the open circle indicates the component $F^{-i}_a$ of the
 gluon field strength tensor.}
 \label{f:qgq}
\end{figure}

\begin{figure}[b]
 \includegraphics{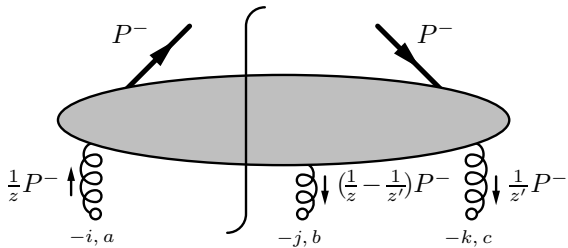}
 \caption{Kinematics of the $ggg$ correlator in Eq.~(\ref{e:ggg}).
 The wiggly lines with the open circles indicate the different components of
 the gluon field strength tensor.} 
 \label{f:ggg}
\end{figure}

Now in order to be more specific we consider essentially two kinds of
correlators in fragmentation: the quark-gluon-quark ($qgq$) correlator as well as
the gluon-gluon-gluon ($ggg$) correlator.
The $qgq$ correlator (see also Fig.~\ref{f:qgq}) is defined by
\begin{eqnarray} \label{e:qgq}
 \Delta_F^{i}(\tfrac{1}{z}, \tfrac{1}{z'})
 &=& \sum_{X} \int \frac{d\xi^+}{2\pi} \, \frac{d\eta^+}{2\pi} \,
     e^{i \tfrac{1}{z'} P^- \xi^+
      + i \big( \tfrac{1}{z} - \tfrac{1}{z'} \big) P^- \eta^+} \nonumber\\
 & & \times \frac{1}{3} \, \text{Tr}\Big[
      \langle 0 | \, g \, t_a F_a^{-i}(\eta^+) \,
      \psi(\xi^+) \, | P, X \rangle \nonumber\\
 & & \times
      \langle P, X | \, \bar{\psi}(0^+) \, | 0 \rangle
     \Big] \,,
\end{eqnarray}
where $v^\pm = (v^0 \pm v^3) / \sqrt{2}$ and $\vec{v}_T = (v^1, v^2)$ are the
light-cone components of a generic 4-vector $v = (v^0, v^1, v^2, v^3)$ and
$P^-$ in Eq.~(\ref{e:qgq}) is the (large) minus momentum of the hadron in the
final state.
In Fig.~\ref{f:qgq} the minus momenta of the three partons are given.
The correlator in Eq.~(\ref{e:qgq}) is a matrix in Dirac space, and the trace
is acting in color space.
We make use of the light-cone gauge $A^- = 0$, in which Wilson lines between
the field operators reduce to the unit matrix.
Note that replacing the quark fields $\psi$ in Eq.~(\ref{e:qgq}) by the
corresponding charge conjugated quark fields $\psi^C$ defines the
$\bar{q}g\bar{q}$ correlator $\bar{\Delta}_F^{i}$.
The $ggg$ correlator containing three field strength tensors (see also
Fig.~\ref{f:ggg}) is given by
\begin{eqnarray} \label{e:ggg}
 \hat{\Gamma}_{F,f}^{i,jk}(\tfrac{1}{z}, \tfrac{1}{z'})
 &=& \sum_{X} \int \frac{d\xi^+}{2\pi} \, \frac{d\eta^+}{2\pi} \,
     e^{i \tfrac{1}{z'} P^- \xi^+
      + i \big( \tfrac{1}{z} - \tfrac{1}{z'} \big) P^- \eta^+} \nonumber\\
 & & \times i f_{abc} \,
      \langle 0 | \, g F_b^{-j}(\eta^+) \,
      F_c^{-k}(\xi^+) \, | P, X \rangle \nonumber\\
 & & \times
      \langle P, X | \, F_a^{-i}(0^+) \, | 0 \rangle \,,
      \vphantom{\Big[}
\end{eqnarray}
with $f_{abc}$ denoting the (antisymmetric) structure constants of the $SU(3)$
group.
If in Eq.~(\ref{f:ggg}) the symmetric structure constants $d_{abc}$ are used
instead of the $if_{abc}$ one obtains the additional, independent
$ggg$ correlator $\hat{\Gamma}_{F,d}^{i,jk}$.

%
%
%
%

We point out that the objects defined above form a complete set of collinear
twist-3 three-parton correlators for fragmentation.
Two other sets can be defined by either substituting the gluon field strength
tensor $F^{-i}$ at position $\eta^-$ with the covariant derivative $D_T^i$ or
by shifting it from the first to the second matrix element in
Eqs.~(\ref{e:qgq}) and~(\ref{e:ggg}).
These sets are, however, not independent of the one we are using.
In the first case they are connected in analogy to Ref.~\cite{eguchi_06a} while
in the second case they are just the Hermitian conjugates of each other.

In the following we will show that the correlators in Eqs.~(\ref{e:qgq})
and~(\ref{e:ggg}) vanish if the longitudinal (minus) momentum of one of the two
partons on the right-hand side~(rhs) of the cut in Figs.~\ref{f:qgq} and~\ref{f:ggg} vanishes.
Precisely these kinematical conditions are what defines the PPMEs.

%
%
{\it II. Support properties of three-parton correlators.}\,---\,We outline our
analysis for the case of the $qgq$ correlator in Eq.~(\ref{e:qgq}).
For this correlator we obtain by inserting a complete set of intermediate
states $| Y \rangle$ and by using the translation properties of field operators
\begin{eqnarray}
 \Delta_F^{i}(\tfrac{1}{z}, \tfrac{1}{z'})
 &=& \sum_{X,Y} \int \frac{d\xi^+}{2\pi} \, \frac{d\eta^+}{2\pi} \,
     e^{i \tfrac{1}{z'} P^- \xi^+
      + i \big( \tfrac{1}{z} - \tfrac{1}{z'} \big) P^- \eta^+} \nonumber\\
 & & \times \frac{1}{3} \, \text{Tr}\Big[
      \langle 0 | \, g \, t_a F_a^{-i}(\eta^+) \, | Y \rangle \nonumber\\
 & & \times
      \langle Y | \, \psi(\xi^+) \, | P, X \rangle \,
      \langle P, X | \, \bar{\psi}(0^+) \, | 0 \rangle
     \Big] \nonumber\\
 &=& \sum_{X,Y} \delta\Big(( \tfrac{1}{z} - 1) P^-
      - {\textstyle\sum_i} p_i^- \Big) \nonumber\\
 & & \times \delta\Big(( \tfrac{1}{z} - \tfrac{1}{z'} ) P^-
      - {\textstyle\sum_j} q_j^- \Big) \nonumber\\
 & & \times \frac{1}{3} \, \text{Tr}\Big[
      \langle 0 | \, g \, t_a F_a^{-i}(0^+) \, | Y \rangle \nonumber\\
 & & \times
      \langle Y | \, \psi(0^+) \, | P, X \rangle \,
      \langle P, X | \, \bar{\psi}(0^+) \, | 0 \rangle
     \Big] \,.\quad
     \label{e:qgq2}
\end{eqnarray}
Here $p_i$ and $q_j$ denote the momenta of the particles in the intermediate
states $| X \rangle$ and $| Y \rangle$, respectively.
As all particles in the intermediate states are on-shell, their momenta have to
satisfy $p_i^2 = m_i^2$ and $p_i^0 \ge 0$ as well as $q_j^2 = m_j^2$ and
$q_j^0 \ge 0$, which leads to $p_i^- \ge 0$ and $q_j^- \ge 0$.
This means, however, that the $\delta$~functions in Eq.~(\ref{e:qgq2}) only
give a nonzero contribution if
\begin{equation}
 \tfrac{1}{z} \ge 1 \quad\text{and}\quad \tfrac{1}{z} \ge \tfrac{1}{z'} \,.
\end{equation}
The case $\tfrac{1}{z} = \tfrac{1}{z'}$ is now of particular interest because
it corresponds to a PPME.
From Eq.~(\ref{e:qgq2}) one finds that here the second $\delta$~function only
contributes if $q_j^- = 0$ for all $j$, which leads to
$q_j^2 = -\vec{q}_{jT}^{\,2} \le 0$.
Combined with the on-shell condition $q_j^2 = m_j^2 \ge 0$ it immediately
follows that all particles in the intermediate state $| Y \rangle$ have to be
massless and move only in the plus direction, i.e., $q_j^- = \vec{q}_{jT}=0$
for all $j$, in order to obtain a nonvanishing contribution to the
$qgq$ correlator.
In this case, however, the matrix element
\begin{equation}
 M^{-i}(q_j) = \langle 0 | \, g \, t_a F_a^{-i}(0^+) \, | Y \rangle
 \label{e:me}
\end{equation}
vanishes.
This matrix element is an antisymmetric Lorentz tensor, which can be expressed
in terms of the Lorentz vectors it depends on\,---\,the momenta $q_j$ and the
polarization vectors $\epsilon(q_j)$ of the massless particles in the
intermediate state\,---\,multiplied by some scalar functions.
One possible decomposition of the matrix element in Eq.~(\ref{e:me}) with
general Lorentz indices is
\begin{eqnarray}
 M^{\mu\nu}(q_j)
 &=& \sum_{m,n} \Big[ q_m^\mu q_n^\nu \, A_{mn}(q_j)
     + q_m^\mu \epsilon^\nu(q_n) \, B_{mn}(q_j) \nonumber\\
 & & + \,\epsilon^\mu(q_m) \, \epsilon^\nu(q_n) \, C_{mn}(q_j)
     - \{ \mu \leftrightarrow \nu \} \Big] \,.
     \label{e:me2}
\end{eqnarray}
However, for massless, on-shell particles moving in the plus direction the
minus components of the polarization vectors $\epsilon(q_j)$ have to vanish
because they correspond to some combination of the unphysical timelike and
longitudinal polarization components.
Therefore, none of the terms in Eq.~(\ref{e:me2}) can contribute to the
matrix element $M^{-i}$ in Eq.~(\ref{e:me}) as we have
$q_j^- = \vec{q}_{jT} = \epsilon^-(q_j) = 0$ for all $j$.
This means that finally the $qgq$ correlator in Eq.~(\ref{e:qgq2}) vanishes for
$\tfrac{1}{z} = \tfrac{1}{z'}$ and we only get a nonzero contribution if
\begin{equation}
 \tfrac{1}{z} \ge 1 \quad\text{and}\quad \tfrac{1}{z} > \tfrac{1}{z'} \,.
 \label{e:sp1}
\end{equation}
On the other hand, if we permute the quark and the gluon field in the first
matrix element in Eq.~(\ref{e:qgq}) before we insert the complete set of
intermediate states $| Y \rangle$ into the $qgq$ correlator, we find
\begin{eqnarray}
 \Delta_F^{i}(\tfrac{1}{z}, \tfrac{1}{z'})
 &=& \sum_{X,Y} \delta\Big(( \tfrac{1}{z} - 1) P^-
      - {\textstyle\sum_i} p_i^- \Big) \nonumber\\
 & & \times \delta\Big( \tfrac{1}{z'} P^-
      - {\textstyle\sum_j} q_j^- \Big) \nonumber\\
 & & \times \frac{1}{3} \, \text{Tr}\Big[
      \langle 0 | \, t_a \, \psi(0^+) \, | Y \rangle \\
 & & \times
      \langle Y | \, g F_a^{-i}(0^+) \, | P, X \rangle \,
      \langle P, X | \, \bar{\psi}(0^+) \, | 0 \rangle
     \Big] \,. \nonumber
\end{eqnarray}
In this case an analysis of the $\delta$~functions analogous to the one for the
$\delta$~functions in Eq.~(\ref{e:qgq2}) yields, that the $qgq$ correlator only
differs from zero if
\begin{equation}
 \tfrac{1}{z} \ge 1 \quad\text{and}\quad \tfrac{1}{z'} > 0 \,.
 \label{e:sp2}
\end{equation}
Here $\tfrac{1}{z'} = 0$, which also corresponds to a PPME, can be excluded
because the matrix element $\langle 0 | \, t_a \, \psi(0^+) \, | Y \rangle$
vanishes if all particles in the intermediate state $| Y \rangle$ are massless.
Therefore, from Eqs.~(\ref{e:sp1}) and~(\ref{e:sp2}) we conclude that the only
nonzero contribution to the $qgq$ correlator in Eq.~(\ref{e:qgq}) comes from
\begin{equation}
 0 \le z \le 1 \quad\text{and}\quad 0 < \tfrac{z}{z'} < 1 \,.
\end{equation}
The same results also hold for the $\bar{q}g\bar{q}$ correlator
and the two $ggg$ correlators.
For the PPMEs this leads to the model-independent results
\begin{eqnarray} \label{e:res_1}
 \Delta_F^{i}(\tfrac{1}{z}, \tfrac{1}{z}) & = &
 \Delta_F^{i}(\tfrac{1}{z}, 0) \, = \, 0 \,,
 \\ \label{e:res_2}
 \bar{\Delta}_F^{i}(\tfrac{1}{z}, \tfrac{1}{z}) & = &
 \bar{\Delta}_F^{i}(\tfrac{1}{z}, 0) \, = \, 0 \,,
 \\ \label{e:res_3}
 \hat{\Gamma}_{F,f/d}^{i,jk}(\tfrac{1}{z}, \tfrac{1}{z}) & = &
 \hat{\Gamma}_{F,f/d}^{i,jk}(\tfrac{1}{z}, 0) \, = \, 0 \,.
\end{eqnarray}
We repeat that the PPMEs for fragmentation, therefore, cannot be responsible
for single spin asymmetries as they all vanish according to
Eqs.~(\ref{e:res_1})--(\ref{e:res_3}).

Note that the support properties of three-parton correlators on the distribution
side can be found by a corresponding analysis.
In that case, however, PPMEs do not vanish which agrees with the
well-established existence of these objects.
The key difference between both analyses is the absence of the vacuum state on
the PDF side.

%
%
{\it III. Universality of transverse-momentum-dependent fragmentation
functions.}\,---\,The vanishing of gluonic pole matrix elements (GPMEs) for
fragmentation is
intimately connected with the universality of transverse-momentum-dependent
two-parton fragmentation functions (TMD FFs)~\cite{boer_03a,bomhof_06}.
To illustrate the relation in more detail we consider the correlator
\begin{eqnarray} \label{e:corr_tmd}
 \Delta^{[\mathcal{U}]}(\tfrac{1}{z}, \vec{k}_T)
 &=& \sum_X \int \frac{d\xi^+}{2\pi} \, \frac{d^2\vec{\xi}_T}{(2\pi)^2} \,
     e^{ik\cdot\xi} \, \frac{1}{3} \,\text{Tr}\Big[
     \langle 0 | \, \mathcal{W}^{[\mathcal{U}]}(0, \xi) \nonumber\\
 & & \times \psi(\xi) \, | P, X \rangle \,
     \langle P, X | \, \bar{\psi}(0) \, | 0 \rangle
     \Big]_{\xi^- = 0} \,,
\end{eqnarray}
which defines TMD FFs when appropriate traces in Dirac space are
taken~\cite{mulders_95,bacchetta_06}.
At leading twist there exist eight FFs.
(We restrict the explicit discussion here to the quark FFs, but our general
conclusions apply to gluon FFs as well.)
The TMD FFs depend both on the longitudinal momentum $k^- = P^-/z$ and the
transverse momentum $\vec{k}_T$ (relative to the detected hadron) of the
fragmenting quark.
The Wilson line $\mathcal{W}^{[\mathcal{U}]}$ cannot be neglected in any
gauge~\cite{ji_02,belitsky_02}, and its path $\mathcal{U}(0; \xi)$ is
determined by the physical process under
consideration~\cite{collins_02,boer_03a,bomhof_04}.
More details about the precise definition of TMD correlators and, in
particular, the choice of Wilson lines can be found
in~\cite{collins_08,hautmann_08,cherednikov_08} and references therein.

The path dependence of the correlator in Eq.~(\ref{e:corr_tmd}) implies a
potential nonuniversality of TMD FFs.
For instance, in semi-inclusive deep inelastic scattering (DIS) one has,
{\it a priori}, past-pointing Wilson lines, while they are future pointing in
$e^+e^-$ annihilation.
In Ref.~\cite{collins_04} it was argued, however, that factorization can be
established such that TMD FFs have the same Wilson lines in both processes.
But the analysis in~\cite{collins_04} made use of a spectator model and was not
carried out to arbitrary order in perturbation theory.
As a consequence, in the community doubts remained concerning the generality of
this result.

In the following we consider $k_T$ moments of the correlator in
Eq.~(\ref{e:corr_tmd}).
We define the zeroth moment by
\begin{equation} \label{e:0_mom}
 \Delta(\tfrac{1}{z}) =
 \int d^2\vec{k}_T \, \Delta^{[\mathcal{U}]} (\tfrac{1}{z}, \vec{k}_T) \,.
\end{equation}
In this case the $k_T$ integration eliminates all the path dependence implying,
in particular, that the three leading twist collinear
FFs~\cite{mulders_95,bacchetta_06}, which are given by $\Delta$ in
Eq.~(\ref{e:0_mom}), are universal.
For the more interesting case of the first $k_T$ moment of the correlator in
Eq.~(\ref{e:corr_tmd}) one finds~\cite{boer_03a,bomhof_06}
\begin{eqnarray} \label{e:1_mom}
 \Delta_{\partial}^{i [\mathcal{U}]}(\tfrac{1}{z})
 &=& \int d^2\vec{k}_T \, k_T^i \, \Delta^{[\mathcal{U}]}
     (\tfrac{1}{z}, \vec{k}_T) \nonumber\\
 &=& \tilde{\Delta}^{i}_\partial(\tfrac{1}{z})
     + C_F^{[\mathcal{U}]} \, \pi \Delta^{i}_F(\tfrac{1}{z},\tfrac{1}{z}) \,.
\end{eqnarray}
Here $\tilde{\Delta}^{i}_\partial$ is the universal part, while the second
term on the rhs in Eq.~(\ref{e:1_mom}) contains all the potential
nonuniversal behavior of the correlator.
The latter is given by a GPME multiplied with the calculable and path-dependent
so-called gluonic pole factor $C_F^{[\mathcal{U}]}$.
According to Eq.~(\ref{e:res_1}) the GPME vanishes, and, therefore, we conclude
that the first moment of the TMD correlator in Eq.~(\ref{e:corr_tmd}) is
universal as well.
This leads to universal second $k_T$ moments of four specific leading twist
TMD FFs, where currently the Collins FF represents the most important one.
This universality is a prerequisite for a combined analysis of
transverse-momentum-dependent data from semi-inclusive DIS and
$e^+e^-$ annihilation~\cite{efremov_06,anselmino_07}, and, in particular, for
the first extraction of the transversity distribution of the
nucleon~\cite{anselmino_07}.

We repeat that universality of TMD FFs was already obtained previously in low
order spectator model calculations~\cite{metz_02,collins_04,yuan_07}.
However, so far no model-independent proof has been given.
Therefore, our result, even though the present analysis is restricted to
certain $k_T$ moments of FFs, provides the first strong model-independent
support for the universality of TMD FFs.

%
%
{\it IV. Conclusions.}\,---\,We have proved in a model-independent way that the
so-called partonic pole matrix elements for fragmentation vanish.
Therefore, in contrast to conjectures in the literature, transverse single spin
asymmetries in hard semi-inclusive reactions are not related to such objects.
Our finding does not mean that single spin asymmetries, in general, cannot be
connected to collinear fragmentation correlators.
This topic requires further investigation.

The vanishing of the partonic pole matrix elements also implies that certain
$k_T$ moments of transverse-momentum-dependent fragmentation functions are
universal.
This result confirms earlier studies, and represents the first fully general
and model-independent proof of this kind.
Hence, we now have additional strong evidence for the universality of
transverse-momentum-dependent fragmentation functions.
For the future one might hope that higher $k_T$ moments of fragmentation
functions can be analyzed along the lines presented in this Letter.

This work has been partially supported by the BMBF (Verbundforschung), the DFG,
and the EU Integrated Infrastructure Initiative Hadronphysics Project under
Contract No.~RII3-CT-2004-506078.

%
%


\begin{thebibliography}{99}

\bibitem{collins_89}
J.\,C.~Collins, D.\,E.~Soper, and G.~Sterman,
Adv.\ Ser.\ Direct.\ High Energy Phys.\ {\bf 5}, 1 (1988).

\bibitem{collins_81c}
J.\,C.~Collins and D.\,E.~Soper,
Nucl.\ Phys.\ B {\bf 194}, 445 (1982).

\bibitem{jaffe_91b}
R.\,L.~Jaffe and X.~Ji,
Nucl.\ Phys.\ B {\bf 375}, 527 (1992).

\bibitem{adams_91a}
D.\,L.~Adams {\it et al.}  (E704 Collaboration),
Phys.\ Lett.\ B {\bf 261}, 201 (1991);
Phys.\ Lett.\ B {\bf 264}, 462 (1991);
Z.\ Phys.\ C {\bf 56}, 181 (1992).

\bibitem{bravar_96}
A.~Bravar {\it et al.}  (E704 Collaboration),
Phys.\ Rev.\ Lett.\ {\bf 77}, 2626 (1996).

\bibitem{adams_03}
J.~Adams {\it et al.}  (STAR Collaboration),
Phys.\ Rev.\ Lett.\ {\bf 92}, 171801 (2004).

\bibitem{adler_05}
S.\,S.~Adler {\it et al.} (PHENIX Collaboration),
Phys.\ Rev.\ Lett.\ {\bf 95}, 202001 (2005).

\bibitem{arsene_08}
I.~Arsene {\it et al.} (BRAHMS Collaboration),
Phys.\ Rev.\ Lett.\  {\bf 101}, 042001 (2008).

\bibitem{abelev_08}
B.\,I.~Abelev {\it et al.} (STAR Collaboration),
Phys.\ Rev.\ Lett.\ {\bf 101}, 222001 (2008).

\bibitem{efremov_81}
A.\,V.~Efremov and O.\,V.~Teryaev,
Sov.\ J.\ Nucl.\ Phys.\ {\bf 36}, 140 (1982)
[Yad.\ Fiz.\ {\bf 36}, 242 (1982)];
Phys.\ Lett.\ B {\bf 150}, 383 (1985).

\bibitem{qiu_91a}
J.-W.~Qiu and G.~Sterman,
Phys.\ Rev.\ Lett.\ {\bf 67}, 2264 (1991);
Nucl.\ Phys.\ B {\bf 378}, 52 (1992);
Phys.\ Rev.\ D {\bf 59}, 014004 (1998).

\bibitem{kouvaris_06}
C.~Kouvaris, J.-W.~Qiu, W.~Vogelsang, and F.~Yuan,
Phys.\ Rev.\ D {\bf 74}, 114013 (2006).

\bibitem{eguchi_06b}
H.~Eguchi, Y.~Koike, and K.~Tanaka,
Nucl.\ Phys.\ B {\bf 763}, 198 (2007).

\bibitem{koike_01}
Y.~Koike,
arXiv:hep-ph/0106260;
AIP Conf.\ Proc.\ {\bf 675}, 449 (2003);
AIP Conf.\ Proc.\ {\bf 675}, 574 (2003).

\bibitem{eguchi_06a}
H.~Eguchi, Y.~Koike, and K.~Tanaka,
Nucl.\ Phys.\ B {\bf 752}, 1 (2006).

\bibitem{bacchetta_05}
A.~Bacchetta, C.\,J.~Bomhof, P.\,J.~Mulders, and F.~Pijlman,
Phys.\ Rev.\ D {\bf 72}, 034030 (2005).

\bibitem{bomhof_06}
C.\,J.~Bomhof and P.\,J.~Mulders,
JHEP {\bf 0702}, 029 (2007).

\bibitem{boer_07a}
D.~Boer, C.\,J.~Bomhof, D.\,S.~Hwang, and P.\,J.~Mulders,
Phys.\ Lett.\ B {\bf 659}, 127 (2008).

\bibitem{metz_02}
A.~Metz,
Phys.\ Lett.\ B {\bf 549}, 139 (2002).

\bibitem{collins_04}
J.\,C.~Collins and A.~Metz,
Phys.\ Rev.\ Lett.\ {\bf 93}, 252001 (2004).

\bibitem{gamberg_08}
L.\,P.~Gamberg, A.~Mukherjee, and P.\,J.~Mulders,
Phys.\ Rev.\ D {\bf 77}, 114026 (2008).

\bibitem{boer_03a}
D.~Boer, P.\,J.~Mulders, and F.~Pijlman,
Nucl.\ Phys.\ B {\bf 667}, 201 (2003).

\bibitem{mulders_95}
P.\,J.~Mulders and R.\,D.~Tangerman,
Nucl.\ Phys.\ B {\bf 461}, 197 (1996) 
[Erratum-ibid.\ B {\bf 484}, 538 (1997)].

\bibitem{bacchetta_06}
A.~Bacchetta {\it et al.},
JHEP {\bf 0702}, 093 (2007).

\bibitem{ji_02}
X.~Ji and F.~Yuan,
Phys.\ Lett.\ B {\bf 543}, 66 (2002).

\bibitem{belitsky_02}
A.\,V.~Belitsky, X.~Ji, and F.~Yuan,
Nucl.\ Phys.\ B {\bf 656}, 165 (2003).

\bibitem{collins_02}
J.\,C.~Collins,
Phys.\ Lett.\ B {\bf 536}, 43 (2002).

\bibitem{bomhof_04}
C.\,J.~Bomhof, P.\,J.~Mulders, and F.~Pijlman,
Phys.\ Lett.\ B {\bf 596}, 277 (2004).

\bibitem{collins_08}
J.\,C.~Collins,
arXiv:0808.2665 [hep-ph].

\bibitem{hautmann_08}
F.~Hautmann and H.~Jung,
AIP Conf.\ Proc.\ {\bf 1056}, 79 (2008).

\bibitem{cherednikov_08}
I.\,O.~Cherednikov and N.\,G.~Stefanis,
arXiv:0811.4357 [hep-ph].

\bibitem{efremov_06}
A.\,V.~Efremov, K.~Goeke, and P.~Schweitzer,
Phys.\ Rev.\ D {\bf 73}, 094025 (2006).

\bibitem{anselmino_07}
M.~Anselmino {\it et al.},
Phys.\ Rev.\ D {\bf 75}, 054032 (2007).

\bibitem{yuan_07}
F.~Yuan,
Phys.\ Rev.\ Lett.\ {\bf 100}, 032003 (2008);
Phys.\ Rev.\ D {\bf 77}, 074019 (2008).

\end{thebibliography}
\end{document}